\documentclass[10pt,journal,final]{IEEEtran}
\usepackage{amsthm,amsmath}
\IEEEoverridecommandlockouts
\usepackage{graphicx, epsfig, epsf}
\usepackage[dvips]{color}   
\usepackage{mathrsfs}
\usepackage{amssymb}
\usepackage[T1]{fontenc}
\usepackage{cite}

\DeclareMathAlphabet{\mathbit}{OT1}{cmr}{bx}{it}

\begin{document}

\title{Joint Design of Channel and Network Coding for Star Networks}

\author{Christian~Koller~\IEEEmembership{Student~Member,~IEEE},
        Martin~Haenggi~\IEEEmembership{Senior~Member,~IEEE},
        J{\"o}rg~Kliewer \IEEEmembership{Senior~Member,~IEEE},
        Daniel~J.~Costello,~Jr.~\IEEEmembership{Life~Fellow,~IEEE}

\thanks{This work was partly supported by NSF grants \mbox{CCF08-30651} and 
\mbox{CCF08-30666}.}
\thanks{This paper was presented in part at the IEEE Information Theory Workshop, Paraty, Brazil, Oct. 2011.}
\thanks{C.~Koller, M.~Haenggi, and D.~J.~Costello,~Jr. are with the Department of Electrical Engineering, University of Notre Dame, Notre Dame, IN 46556 USA (e-mail: ckoller@nd.edu; mhaenggi@nd.edu; dcostel1@nd.edu).}
\thanks{J. Kliewer is with the Klipsch School of Electrical and Computer Engineering,
New Mexico State University, Las Cruces, NM 88003-8001 USA (e-mail: jkliewer@
nmsu.edu).}
} 

\maketitle

\vspace{-5ex}
\begin{abstract}

Channel coding alone is not sufficient to reliably transmit a message of finite length $K$ from a source to one or more destinations as in, e.g., file transfer. To ensure that no data is lost, 
it must be combined with rateless erasure correcting schemes on a higher layer, such as a time-division multiple access (TDMA) system paired with automatic repeat request (ARQ) or random linear network coding (RLNC).
We consider binary channel coding on a binary symmetric channel (BSC) and $q$-ary RLNC for erasure correction in a star network, where $Y$ sources send messages to each other with the help of a central relay. In this scenario RLNC has been shown to have a throughput advantage over TDMA schemes as $K\rightarrow\infty$ and $q\rightarrow\infty$.
In this paper we focus on finite block lengths and compare the expected throughputs of RLNC and TDMA. 
For a total message length of $K$ bits, which can be subdivided into blocks of smaller size prior to channel coding, we obtain the channel coding rate and the number of blocks that maximize the expected throughput of both RLNC and TDMA, and we find that TDMA is more throughput-efficient for small message lengths $K$ and small $q$.

\end{abstract}

\section{Introduction}

Random linear network coding (RLNC) has recently been shown to improve network 
performance in several broadcast and multicast scenarios.
For example, considering packet erasure channels on the link layer, RLNC is known to improve throughput and reduce delay for wireless broadcast 
\cite{LMKE08, AOMA08, GTK07, SE07}. Further, in \cite{SE07} the joint design of network coding and medium access control protocols was considered.

In contrast to the above work, we consider the joint design of channel and network coding.
We assume that the size of a block is not predetermined and, for a finite message length $K$, the sources in a network may choose the number of data blocks so that the throughput of the overall system is maximized.

The joint design and optimum rate allocation between channel and network coding 
for the block fading channel has been investigated in \cite{CHK09, BZWWP08, CW11}, where the tradeoff between the two schemes is analyzed as the block length on the physical layer gets large and the probability of block erasure is given by the outage probability of the block fading channel, under the assumption that the coherence time of the fading channel grows with the block length.

Joint error and erasure correcting coding for finite message lengths was analyzed in \cite{VM05, Xiao10, itw11}.
In \cite{VM05} the authors bound the performance of random coding on the physical and link layer using error exponents to trade off system throughput and delay.
In \cite{Xiao10} the combination of RLNC and continuous-time orthogonal waveform 
channels was investigated. Both papers aim to maximize throughput given a maximum delay constraint. By contrast, in this paper we do not enforce a maximum delay constraint, but focus instead on the expected throughput for reliable communication, assuming the senders continue to transmit until the receivers have correctly received the entire message as in, e.g., file transfer.
Thus we use the expected throughput of the network as the performance metric and compare it to a TDMA system using ARQ.

\begin{figure}[t] 
    \centering
		\includegraphics[width=0.95\columnwidth]{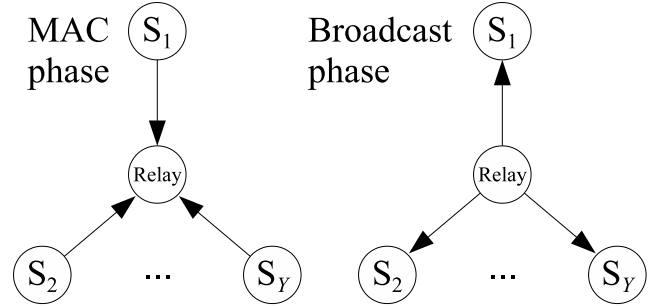}
		\caption{Star network in which $Y$ sources communicate over noisy BSCs with the help of a central relay.}
		\label{fig:Star}
\end{figure}

We consider a star network as depicted in Fig.~\ref{fig:Star}. With the help of a central
relay, $Y$ sources, $S_1,\ldots,S_Y$, communicate with each other over noisy binary symmetric channels (BSCs).
We assume there is no direct path between any of the sources, i.e., they
are only connected to the central relay, which receives transmissions from all
sources and can broadcast to all sources.
We consider the case where each source $S_i$ has a message of finite length $K$ bits
that is intended for all the other $Y-1$ sources $S_j$, $j=1,\ldots,Y$, $j\neq i$.

In this setting, channel coding alone is not sufficient to guarantee reliable communication.
To ensure that no data is lost, channel coding on the physical layer must be combined  with rateless erasure correcting schemes, such as a time-division multiple access (TDMA) system paired with automatic repeat request (ARQ) \cite{LCM84} or $q$-ary RLNC \cite{ACLY00}, where RLNC was shown to be asymptotically optimal, as $K\rightarrow \infty$ and $q \rightarrow \infty$, in \cite{DGPHE06}.

We define the time that it takes to transmit one bit as a time unit and, when maximizing the expected throughput, we minimize the expected number of time units it takes to successfully transmit $Y$ messages from $Y$ sources to the other $Y-1$ sources.

More specifically, we aim to answer the questions:
\begin{itemize}
	\item Given RLNC over GF($q$) and a message of length $K$ bits at each source, what is the number of blocks $m$ that the sources should use to transmit so that the expected system throughput is maximized?
	\item What is the channel coding rate for each individual block that maximizes
	system throughput?
	\item How does the throughput of RLNC compare to the throughput of TDMA as a function of the number of blocks $m$ and the Galois field size $q$?
\end{itemize}
Our goal is to jointly find the number of blocks and the channel coding rate that maximizes system throughput.
Choosing a star network as a model allows us to combine several prominent features of more general networks. For example, for $Y=2$ sources, the star network reduces to a two-hop line network with a relay where the two ends communicate with each other.
Additionally, in the RLNC case, the star network model includes a multiple-access channel (MAC) phase, where all the sources simultaneously transmit to a central relay, followed by a broadcast phase, where the relay transmits to all sources, as illustrated in Fig.~\ref{fig:Star}. We first analyze these two phases separately before combining them to maximize the throughput of the star network. 

In our analysis, we take the coding overhead of RLNC into account.
Similar to other rateless coding schemes \cite{Lub02, Sho06}, 
RLNC over a finite number of blocks $m$ and GF($q$) exhibits a coding overhead, i.e, a receiver on average needs to correctly receive more than $m$ blocks to be able to decode.
Note that the coding overhead is a property of the code itself and is
different from the signaling overhead, which is usually appended to the data in a block header.

\section{System Model}

\subsection{Star Network Setup}

We consider a star network where $Y$ sources $S_1,\ldots,S_Y$ 
communicate with each other with the help of a central relay as shown in
Fig.~\ref{fig:Star}.

Using RLNC, data transmission is divided into two phases, the MAC phase, 
where all the sources simultaneously transmit to 
the relay, and the broadcast phase, where the central relay transmits to 
all sources.
\begin{figure}[t] 
    \centering
		\includegraphics[width=0.95\columnwidth]{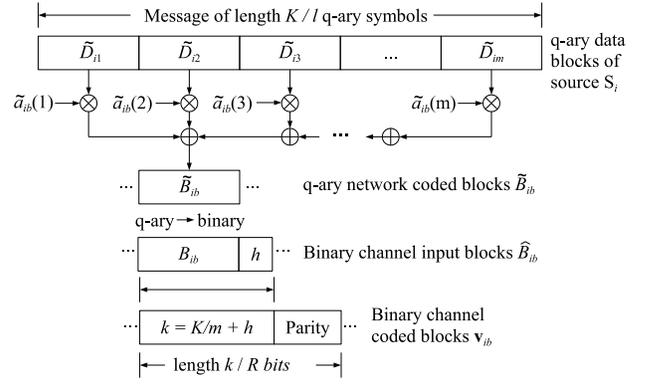}
		\caption{A source $S_i$ divides its message of length $K$ bits into $m$ blocks. RLNC over GF($q$) is then used to create $q$-ary network coded blocks $\tilde{B}_{ib}$. After q-ary to binary conversion, a header of size $h$ bits is appended to each block and the resulting block of size $k=K/m+h$ bits is protected by a linear channel code of rate $R$ to create binary channel coded blocks $\mathbf{v}_{ib}$.}
		\label{fig:Blocks}
\end{figure}

As shown in Fig.~\ref{fig:Blocks}, a source $S_i$, $i=1,\ldots,Y$, splits its message of length $K$ bits into $m$ binary data blocks $D_{ij}$, $j=1,\ldots,m$, of length $K/m$ bits, or equivalently $m$ $q$-ary data blocks $\tilde{D}_{ij}$, $j=1,\ldots,m$, of length $K/(ml)$ $q$-ary symbols. We assume that $q$ is a power of two, i.e., $q=2^l$ and that $K$ is divisible by $ml$.

A source $S_i$ then performs RLNC on its $m$ data blocks to create a network coded block $\tilde{B}_{ib}$ by choosing a vector $\tilde{\mathbf{a}}_{ib}$ of length $m$ of coefficients from GF($q$), where the index $b$ does not have a fixed range, since as many blocks are created as are necessary to achieve reliable communication.
The coded block $\tilde{B}_{ib}$ is then the linear combination of the $m$ data blocks multiplied by the corresponding components of the coefficient vector $\tilde{\mathbf{a}}_{ib}$, i.e.,
\begin{equation} \label{eq:RLNCencoding}
		\tilde{B}_{ib} = \sum_{j=1}^m \tilde{a}_{ib}(j)\tilde{D}_{ij},
\end{equation}
which can also be represented as a binary block of length $K/m$ bits using the notation $B_{ib}$.

A header of constant size $h$ bits is then appended to each coded block $B_{ib}$ to form a channel input block $\hat{B}_{ib}$ of length $k=K/m+h$ bits. The header can, for example, contain a cyclic redundancy check (CRC) to detect decoding failures. Finally, each channel input block $\hat{B}_{ib}$ is protected by a binary channel code of rate $R$, forming the channel coded block $\mathbf{v}_{ib}$.

\paragraph{MAC phase} During the MAC phase, all sources transmit to the relay simultaneously.
We model the channel from the sources to the relay as a binary adder channel
\cite{KS64, GW75, KL76}, so that the relay receives a value equal to the (real) sum of the bits sent by the sources plus a noise term\footnote{For simplicity, the binary adder channel model assumes that the sources transmit on-off pulses, the relay accumulates the total energy received from all sources in each time slot, and a common clock synchronizes all sources.}. The relay then quantizes each received value to the nearest integer and makes a hard decision. If the quantized value is even, it decides a received zero, and if the quantized value is odd, it decides a received one, so that the resulting received bit can be modeled as the modulo-2 superposition of the bits sent by all the sources plus a noise bit.
Equivalently, the received superimposed vector at the relay is given by
\begin{equation}
	\mathbf{r}_{b} = \mathbf{v}_{b} \oplus \mathbf{e} = 
	\mathbf{v}_{1b} \oplus \mathbf{v}_{2b} \oplus \ldots \oplus 
					\mathbf{v}_{Yb} \oplus \mathbf{e},
\end{equation}
where $\oplus$ symbolizes modulo-2 addition and $\mathbf{e}$ is a binary vector whose
elements are Bernoulli i.i.d. random variables with probability $p_\mathrm{mac}$.
This is an extension of the well known two-user binary adder channel model and is equivalent to sending $\mathbf{v}_{1b} \oplus \mathbf{v}_{2b} \oplus \ldots \oplus \mathbf{v}_{Yb}$
over an i.i.d. memoryless binary symmetric channel (BSC) with crossover probability $p_\mathrm{mac}$.
Since we are using linear codes, the modulo-2 superposition of valid codewords results again in a valid codeword $\mathbf{v}_{b}$, which the relay attempts to decode.
If the relay is able to decode, it broadcasts $\mathbf{v}_{b}$
to the sources. Note that the relay does not perform any network coding; 
it only decodes the superposition of the channel coded blocks from the sources.
Should the relay not be able to decode, it does not transmit.
We assume the sources can sense the channel, so if the relay fails to decode and does not transmit, the sources immediately transmit another channel coded block and we have another MAC phase.

\paragraph{Broadcast phase} During the broadcast phase we assume that the relay is connected to each of the destinations via independent BSCs with crossover probability $p_\mathrm{br}$.
Note that, since we are considering finite block lengths, independent BSCs can lead to some sources being able to decode the message from the relay, while others fail to do so. We assume that the sources are at about the same distance from the relay, and thus experience the same path loss, so that they share a common channel crossover probability.\footnote{The analysis would still be possible, but more tedious, if the channels had different crossover probabilities.}

Each channel coded block $\mathbf{v}_{b}$ sent by the relay during the broadcast phase is a linear combination of $Ym$ data blocks, multiplied by a corresponding set of $Ym$ network coding coefficients.
To be able to decode, a source must know the network coding coefficients $\tilde{\mathbf{a}}_b$ that were used to create each superimposed block sent by the relay. One method of letting the receivers know $\tilde{\mathbf{a}}_b$ is to add the coefficients to the header information. Another way, which we adopt in this paper, is to assume that the sources and the receivers use $Y$ synchronized pseudo-random number generators, each source with a different seed, that generate the sequences for $\tilde{\mathbf{a}}_b$.

The column vector of $Ym$ network coding coefficients $\tilde{\mathbf{a}}_{b} = [\mathbf{a}_{1b}, \ldots, \mathbf{a}_{Yb}]'$ corresponding to a block $b$ is the $b$th column in the generator matrix $\mathbf{G}$ employed by the RLNC in the star network, and $\tilde{B}_b=\sum_{i=1}^{Y}\tilde{B}_{ib}$, the superposition of the network coded blocks, can be viewed as a code symbol of the RLNC.

When a source $S_i$ receives a superimposed channel coded block $\mathbf{v}_{b}$ from the relay, it first decodes the binary channel code to obtain the modulo-2 superposition of the channel input blocks $\hat{B}_b=\sum_{i=1}^{Y}\hat{B}_{ib}$. If decoding is successful, as indicated by the CRC in the header, the header of size $h$ bits is removed and, after binary to $q$-ary conversion, the $q$-ary superposition of the network coded blocks $\tilde{B}_b$ is obtained. Source $S_i$ then subtracts its own contribution to $\tilde{B}_b$, which is $\tilde{B}_{ib}$, 
and stores the superposition of the other $Y-1$ network coded blocks $\tilde{B}_{jb}$, $j=1,\ldots,Y$, $j\neq i$, as an element in a vector of received RLNC symbols. It also stores the subset of $(Y-1)m$ network coding coefficients in $\tilde{\mathbf{a}}_{b}$ involved in creating the superposition $\tilde{B}_{jb}$ as a column in its coefficient matrix $\mathbf{G}_i$, the perceived generator matrix of the RLNC from the point of view of source $S_i$. Note that the rows and columns of $\mathbf{G}_i$, a subset of the matrix $\mathbf{G}$, do not contain information about blocks that were not correctly received by the relay or by source $S_i$.

Once a source $S_i$ has received enough blocks from the relay to form a matrix $\mathbf{G}_i$ with $(Y-1)m$ linearly independent columns, it can recover the $(Y-1)m$ data blocks from the other sources by inverting the matrix $\mathbf{G}_i$ and multiplying it by its vector of received RLNC symbols.
On average, a source $S_i$ needs to collect more than $(Y-1)m$ correctly received blocks to form a $\mathbf{G}_i$ of rank $(Y-1)m$, and in Subsection~\ref{sub:RLNC} we bound the expected overhead of RLNC for finite size Galois fields.

Once a source $S_i$ has collected enough blocks to decode the RLNC, it sends a single acknowledgment (ACK) to the relay. Once the relay has collected $Y$ ACKs from the $Y$ sources, it broadcasts an ACK to the sources, terminating transmission. All sources continue to transmit until they receive an ACK from the relay. We assume that the transmission of an ACK is instantaneous and reliable, i.e., that it does not consume any resources and it is never received erroneously.\footnote{We assume the length of the ACK is negligible compared to the length of the message and that it is protected by a more powerful error-correcting code than the message itself.}

As a reference scheme we consider TDMA transmission of the sources, paired with ARQ. 
We also assume a source splits its message into $m$ data blocks, but no network coding is used. 
The MAC phase in Fig.~\ref{fig:Star} is replaced by a TDMA phase, where only one source transmits to the relay at a given time and the individual data blocks are again protected by a binary channel code of rate $R$. The transmitting source $S_i$ repeats the transmission of a channel coded block as many times as is necessary for the relay to receive the data block correctly, at which point the relay transmits an ACK.
After the relay has received the data block correctly it broadcasts it to all sources. 
When a source receives the data block correctly, it sends an ACK to the relay.
The relay repeats the broadcast transmission as many times as is necessary until all $Y-1$ sources $S_j$, $i=1,\ldots,Y$ and $j \neq i$, receive the data block correctly. 
After the steps described above have been successfully completed for source $S_i$, it is the turn of the next source to transmit a data block to the relay, and the sources are scheduled in a round robin fashion with $m$ rounds. After each source has successfully transmitted $m$ data blocks, the transmission ends.

\subsection{A Motivating Example}

Consider the case where the error probability on all BSC links is zero, i.e., $p_\mathrm{mac}=p_\mathrm{br}=0$, and RLNC is performed over an infinitely large Galois field. Furthermore, let $m=1$ and the header size $h=0$.
Each of the $Y$ sources has a message, e.g., a file of size $K$ bits to transmit 
to the others. Using the TDMA scheme, for every one of the $Y$ sources, there is a phase where the source transmits $K$ bits to the relay followed by a phase where the relay broadcasts $K$ bits.
The average throughput of the TDMA scheme is thus given by
\begin{equation*}
	{T}_{\mathrm{TDMA}} = \frac{1}{2}.
\end{equation*}
For the RLNC scheme, an individual source must collect $Y-1$ blocks of 
$K$ bits in order to be able to decode, and the throughput is given by
\begin{equation*}
	{T}_{\mathrm{RLNC}} = \frac{YK}{2(Y-1)K}.
\end{equation*}
The RLNC scheme thus achieves a throughput gain of
\begin{equation} \label{eq:MotEx}
	\frac{{T}_{\mathrm{RLNC}}}{{T}_{\mathrm{TDMA}}} = \frac{Y}{Y-1}
\end{equation}
over the TDMA scheme. The gain of RLNC is largest when only 2 nodes exchange 
information and decreases to one as the number of nodes in the star network gets large. We now describe the channel and network coding in more detail.

\subsection{Channel Coding}

We consider random coding on the physical layer and use two different
approaches to bound the performance of channel coding.

\begin{enumerate}
\item
The block error probability $\epsilon$ of random coding on the BSC with a code rate $R$ can be bounded using the random coding error exponent ${E}(R)$:
\begin{equation} \label{eq:epsilon}
		\epsilon \leq 2^{-n\, {E}(R)},
\end{equation}
where $n=k/R$ is the block length of the code and $k = K/m + h$ bits.
Using the union bound, the random coding error exponent for the BSC is given by \cite{BF02}
\begin{equation} \label{eq:ErrorExp}
					{E}(R) = R_0-R,
\end{equation}
where $R_0$, the cutoff rate of the channel, depends on the crossover probability $p$ of the
BSC and is given by
\begin{equation*}
					R_0 = -\log_2 \left( \frac{1}{2}+\sqrt{p(1-p)} \right).
\end{equation*}
Above the so-called critical rate $R_{crit}$, a tighter upper bound on the block error probability is obtained by using the sphere packing exponent. However, the union bound is often used to approximate the performance of codes of practical length, and hence we adopt the simple form of \eqref{eq:ErrorExp}, which also allows us to obtain analytical expressions for the optimum channel coding rate and optimum number of data blocks. 
We use the above method to bound the performance of channel coding in Sections~\ref{sec:MAC}--\ref{sec:joint}.

\item
Tighter bounds on the achievable channel coding rate given a block error 
probability $\epsilon$ have been derived in \cite{PPV10}, in the following referred to as the PPV bound.
The relationship between the achievable code rate $R$, the error probability $\epsilon$, the
length of the channel code $n=k/R$, and the BSC crossover probability $p$ can be written as
\begin {equation} \label{eq:Polyanskiy}
	R = C - \sqrt{\frac{p(1-p)}{n}}\log_2\left( \frac{1-p}{p} \right) Q^{-1}(\epsilon) + \frac{\log_2(n)}{2n},
\end{equation}
where
\begin{equation*}
	C = 1 - \mathbb{H}(p)
\end{equation*}
is the channel capacity of the BSC, $\mathbb{H}(x) = -x\,\log_2(x)-(1-x)\log_2(1-x)$
is the binary entropy function,
\begin{equation*}
Q(x) = \frac{1}{\sqrt{2\pi}} \int_{x}^{\infty} e^{-u^2/2} du
\end{equation*}
is the tail probability of the Gaussian distribution, and $Q^{-1}(x)$ is its inverse.
In contrast to the first approach described above, using the PPV bound
allows us to consider code rates up to the channel capacity.
We use the above method to bound the performance of channel coding in
Section~\ref{sec:Poly} and compare the results to those obtained in Sections~\ref{sec:MAC}--\ref{sec:joint}.
\end{enumerate}

\subsection{The Expected Overhead of Random Linear Network Coding}
\label{sub:RLNC}

In this subsection we bound the expected coding overhead of RLNC in the star network.
As depicted in Fig.~\ref{fig:Blocks}, each source constructs a random linear network code over $m$ data blocks before sending a network coded block to the relay.
Considering a single source on its own and RLNC over GF($q$), the probability that $m+x$ independently created column vectors of network coding coefficients $\tilde{\mathbf{a}}$ form an $m \times (m+x)$ matrix of rank $m$, i.e.,
the probability that $m+x$ network coded blocks are sufficient to decode the RLNC of that source is given by \cite{LPC10}
\begin{equation} \label{eq:Psucc}
		{P}_\mathrm{success}(m,x,q) = \prod_{i=1}^m \left( 1-q^{-x-i} \right).
\end{equation}

\begin{figure}[t]
    \centering
		\includegraphics[width=0.95\columnwidth]{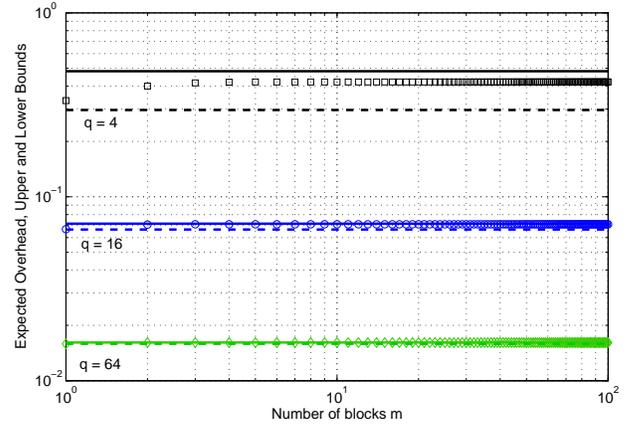}
		\caption{Expected overhead $X(m,q)$ (markers) in blocks with the upper (solid lines) and lower (dashed lines) bounds on the expected overhead in blocks.}
		\label{fig:ULBoundXq1}
\end{figure}
In the star network, a block broadcast by the relay is a linear combination of $Ym$ data blocks and every source can reduce the problem of decoding the network code to that of decoding the $(Y-1)m$ unknown data blocks by subtracting out its own data.
Then, since the network coding coefficients are chosen independently at all sources, the probability that all $Y$ sources can construct an invertible matrix of rank $(Y-1)m$ from $(Y-1)m+x$ correctly received blocks is given by
\begin{equation} \label{eq:PsuccN1}
\begin{aligned}
		{P}^*_\mathrm{success}(m,x,q,Y) = & \left( {P}_\mathrm{success}((Y-1)m,x,q) \right)^Y\\
			 = & \left( \prod_{i=1}^{(Y-1)m} \left( 1-q^{-x-i} \right) \right)^Y.
\end{aligned}
\end{equation}
We now use a result from \cite{LPC10} to bound \eqref{eq:Psucc} as
\begin{equation} \label{eq:boundsPS}
	1-\frac{1}{q-1} q^{-x} < {P}_\mathrm{success}(m,x,q) \leq 1-q^{-x-1},
\end{equation}
which can be used to derive upper and lower bounds on the expected overhead of 
RLNC in the star network that are independent of the number of data blocks $m$.
Using \eqref{eq:PsuccN1} and \eqref{eq:boundsPS}, the probability ${P}^*(m,x=i,q,Y)$ that overhead $x=i$ blocks is required to decode in the star network is upper bounded by
\begin{align}
		{P}^*(m,&x=i,q,Y) = {P}^*_\mathrm{success}(m,i,q,Y) \nonumber\\
						& - {P}^*_\mathrm{success}(m,i-1,q,Y) \nonumber\\
					< & \left( 1-q^{-i-1} \right)^Y - 
							\left( 1-\frac{q^{-i+1}}{q-1}  \right)^Y\nonumber\\
					= & \sum_{j=1}^Y \binom{Y}{j} (-1)^{j+1} 
							\left( \frac{q^j}{(q-1)^j} - \frac{1}{q^j} \right) q^{-ji}.
\end{align}
The expected coding overhead $X^*(m,q,Y)$ of RLNC in blocks in the star network is thus upper bounded by
\begin{equation} \label{eq:ubXq}
		\begin{aligned}
			X^*(m&,q,Y) = \sum_{i=1}^\infty i {P}^*(m,x=i,q,Y)\\
					< & \sum_{j=1}^Y \binom{Y}{j} (-1)^{j+1} 
							\left( \frac{q^j}{(q-1)^j} - \frac{1}{q^j} \right)
							\sum_{i=1}^\infty i\, q^{-ji}\\
					= & \sum_{j=1}^Y \binom{Y}{k} (-1)^{j+1}
							\frac{q^{2j}-(q-1)^j} {(q-1)^j(q^j-1)^2} \triangleq X^*(q,Y).
		\end{aligned}
\end{equation}
In the same way, we can lower bound the expected overhead of RLNC in blocks as
\begin{equation} \label{eq:lbXq}
			X^*(m,q,Y) > \sum_{j=1}^Y \binom{Y}{j} (-1)^{j+1}
							\frac{ (q^2-q)^j-q^j }{ (q-1)^j(q^j-1)^2 }.
\end{equation}
Both bounds \eqref{eq:ubXq} and \eqref{eq:lbXq} are independent of the number of data
blocks $m$ and tend to zero as the size of the Galois field gets large.

Fig.~\ref{fig:ULBoundXq1} shows the actual expected overhead for RLNC of a single source for several 
Galois field sizes $q$ compared to the upper bound \eqref{eq:ubXq} for $Y=1$, 
displayed as solid lines, and the lower bound \eqref{eq:lbXq}, displayed as dashed lines,
where the expected overhead $X(m,q)$ of RLNC in blocks is given by \cite{LMS09}
\begin{equation} \label{eq:Xq}
	X(m,q) = \sum_{i=1}^m \frac{1}{q^i-1}.
\end{equation}
Although \eqref{eq:Xq} is not independent of the number of data blocks $m$, 
Fig.~\ref{fig:ULBoundXq1} shows that the expected overhead in blocks is well approximated by a constant fractional number of blocks. The larger the Galois field size $q$, the better the performance of RLNC, and the quicker the expected overhead converges to a constant. As $q$ increases, the upper and lower bounds on the expected overhead become tighter, and for $q=64$ they are almost indistinguishable.

\begin{figure}[t] 
    \centering
		\includegraphics[width=0.95\columnwidth]{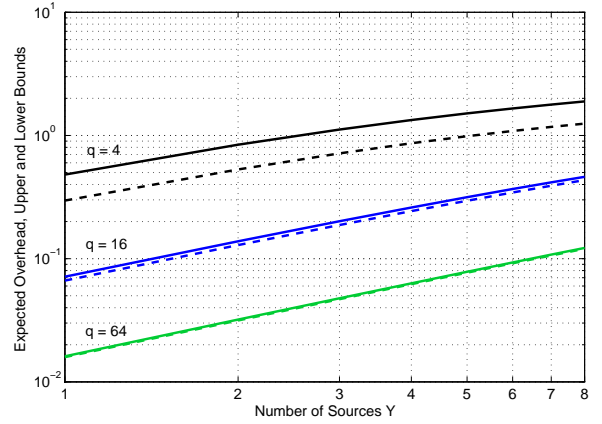}
		\caption{Upper (solid lines) and lower (dashed lines) bound on the expected overhead
		of RLNC in blocks for different numbers of sources $Y$.}
		\vspace{-2ex}
		\label{fig:ULBoundXqN}
\end{figure}
Fig.~\ref{fig:ULBoundXqN} shows the upper \eqref{eq:ubXq} and lower \eqref{eq:lbXq} bounds
on the expected overhead of RLNC in blocks for different numbers of sources $Y$. As the number of 
sources $Y$ increases, the expected overhead of RLNC increases as well.

Modeling the expected coding overhead of RLNC as a constant fractional number of blocks leads to opposing optimization criteria for channel coding and RLNC when a message of finite size $K$ bits is divided into $m$ data blocks:
\begin{itemize}
	\item More data blocks, and thus shorter channel coded blocks, lead to a smaller coding overhead of RLNC in bits.
	\item Longer channel coded blocks, and thus fewer data blocks, lead to more
	powerful channel codes.
\end{itemize}

In the following, when investigating the optimum number of data blocks $m$ and the optimum channel coding rate $R$, we first consider the MAC phase and the broadcast phase separately before finding the values that jointly maximize throughput for the star network.

\section{The MAC Phase} \label{sec:MAC}

In this section we optimize the throughput for the MAC phase and do not consider the broadcast phase in the optimization.
To this end, assume that the channels from the relay to the sources are error-free, i.e., $p_\mathrm{br}=0$, so that the relay does not need a channel code, and that the relay removes the $h$ header bits prior to broadcasting. (When $p_\mathrm{br}=0$, appending a CRC to detect decoding failures is not necessary.)

Using random coding error exponents, we obtain the channel coding rate $R$ and the number of data blocks $m$ that minimize the expected number of transmissions at the sources and thus maximize the throughput. We then compare the results for RLNC to the optimum rate and number of data blocks for TDMA.

Modeling the expected coding overhead of RLNC as a constant fractional number of blocks \eqref{eq:ubXq}, on average each source must collect $(Y-1)m + X^*(q,Y)$ network coded blocks to be able to decode, and the expected number of channel coded 
blocks that the sources need must transmit is thus given by
\begin{equation} \label{eq:EMuni}
	{M}^{\mathrm{mac}}_{\mathrm{RLNC}} \approx \frac{(Y-1)m + X^*(q,Y)}
					{1-\epsilon_\mathrm{mac}},
\end{equation}
where $\epsilon_\mathrm{mac}$ is the block error rate of channel coding for a BSC with crossover probability $p_\mathrm{mac}$.
Using the union bound random coding error exponent to approximate the block erasure rate \eqref{eq:epsilon} and letting $n=k/R=(K/m+h)/R$ be the size of a channel coded block in bits, we obtain from \eqref{eq:EMuni}
\begin{equation} \label{eq:ENmac}
	{N}^{\mathrm{mac}}_{\mathrm{RLNC}} \approx \frac{k\left((Y-1)m + X^*(q,Y)\right)}
					{R\left( 1- 2^{ -k \left( R_0/R -1 \right)} \right)}
\end{equation}
for the expected number of bits that must be sent by the sources.
To minimize the expected number of bits sent, i.e., to maximize throughput, we use the partial derivatives of \eqref{eq:ENmac} with respect to $R$ and $m$ to find the optimum channel coding rate and the optimum number of data blocks, respectively.

For TDMA, a total of $Ym$ blocks must be transmitted to the relay by the $Y$ sources and we have
\begin{equation} \label{eq:EMtdma}
{M}^\mathrm{mac}_{\mathrm{TDMA}} \leq \frac{Ym}
					{1-\epsilon_{mac}}.
\end{equation}
Using equations \eqref{eq:EMtdma} and \eqref{eq:epsilon} for transmission over a BSC with crossover probability $p_\mathrm{mac}$, we obtain
\begin{equation} \label{eq:ENtdma}
	{N}^\mathrm{mac}_{\mathrm{TDMA}} \leq \frac{Y(K+mh)}
					{R\left( 1-2^{ -\left(\frac{K}{m} + h\right)
					\left( \frac{R_0}{R} -1 \right)} \right)}
\end{equation}
for the expected total number of transmitted bits.

\subsection{The Optimum Channel Coding Rate}

\begin{figure}[t] 
    \centering
		\includegraphics[width=0.95\columnwidth]{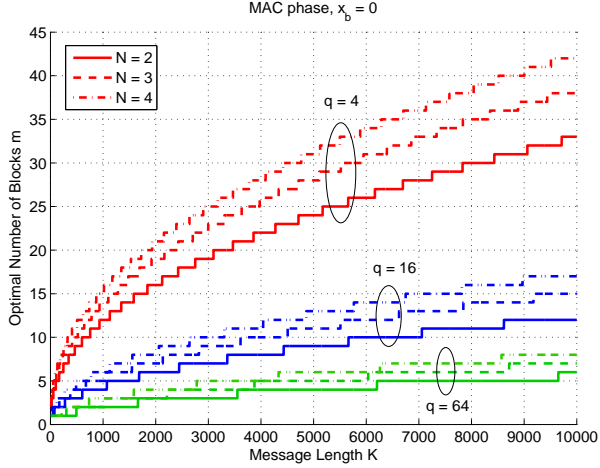}
		\caption{Optimum number of blocks $m$ given the message length $K$, the number of sources $Y$, and header size $h=0$ for RLNC over different Galois field sizes $q$.}
		\label{fig:OptM1}
\end{figure}
Taking the partial derivative of \eqref{eq:ENmac} with respect to $R$ and setting it to zero,
we obtain
\begin{equation*} \label{eq:diff1}
					1-2^{ -k \left( \frac{R_0}{R} -1 \right) } - 
					\ln(2)k\frac{R_0}{R} 2^{ -k \left( \frac{R_0}{R} -1 \right) } = 0,
\end{equation*}
where $k=nR=K/m+h$ is the block length before channel coding.
Using the substitution $t=\ln(2)k\frac{R_0}{R}$, we then obtain
\begin{equation*}
	 -(t+1)e^{-(t+1)} = -e^{-{\ln(2)k+1}},
\end{equation*}
which can be solved using the Lambert-W function $\textrm{W}(x)$ given by 
\begin{equation*}
	x \equiv \textrm{W}(x) e^{\textrm{W}(x)}.
\end{equation*}
The optimum channel coding rate as a fraction of the cutoff rate of the channel is then given by
\begin{equation} \label{eq:SolRate}
	 \frac{R}{R_0} = \frac{-\ln(2)k}{\textrm{W}_{-1}\left(-e^{-(\ln(2)k+1)} \right) +1 },
\end{equation}
where $\textrm{W}_{-1}(x)$ represents the lower branch of the Lambert-W function \cite{Barry93}.
\footnote{For negative arguments, the Lambert-W function has two solutions. Since the ratio $R/R_0$ must be between zero and one, we require $\textrm{W}(x)\leq -1$, so the solution must be on the lower branch of the Lambert-W function.}
From \eqref{eq:SolRate} we see that the optimum channel coding rate ratio $R/R_0$ is only a function of the block length $k$ and is independent of the expected overhead $X^*(q,Y)$ of RLNC and the number of sources $Y$.
It is thus also the optimum channel coding rate for a scheme employing TDMA.

\begin{figure}[t] 
    \centering
		\includegraphics[width=0.95\columnwidth]{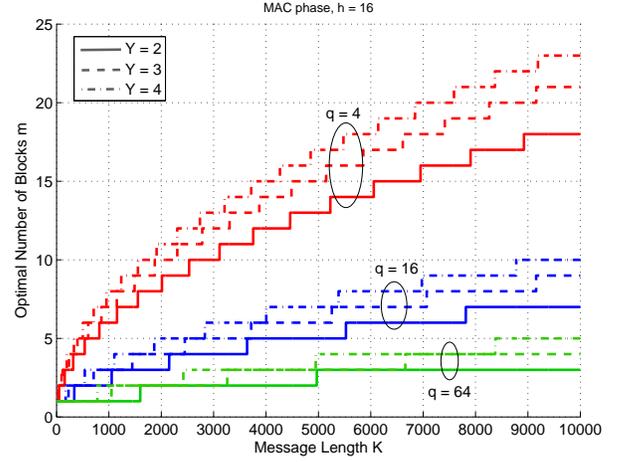}
		\caption{Optimum number of blocks $m$ given the message length $K$, the number of sources $Y$, and header size $h=16$ for RLNC over different Galois field sizes.}
		\label{fig:MoptMACxb16}
\end{figure}
To evaluate the Lambert-W function we use the closed form approximation \cite{Barry93}
\begin{equation*}
	\textrm{W}_{-1}(x) \approx \ln(-x) -\frac{1}{A_1} \left[ 
					1- \frac{1}{ 1+\frac{A_1\sqrt{\sigma/2}}{1-A_2\sigma \exp\{-A_3\sqrt{\sigma} \}} }
	\right],
\end{equation*}
where
\begin{equation*}
	\sigma = -\ln(-x)-1,
\end{equation*} 
$A_1=0.3361$, $A_2=0.0042$, and $A_3 = 0.0201$. The approximation has a maximum relative error of only $0.025\%$. Using the approximation, we see that, as the block length $k$ increases, the optimum channel coding rate ratio $R/R_0$ tends to 1.

\subsection{The Optimum Number of Blocks}

Now taking the partial derivative of \eqref{eq:ENmac} with respect to $m$ and setting it to zero, we obtain
\begin{equation} \label{eq:DerMmac}
\begin{aligned}
		& 2^{ z \left(\frac{K}{m}+h\right) } = \\
		& \left(1 + \frac{\ln(2)zK\left(\frac{K}{m}+h\right)(X^*(q,Y)+m(Y-1))}
									{K\,X^*(q,Y)-h\,m^2(Y-1) } \right),
\end{aligned}
\end{equation}
where $z=(R_0/R)-1$.
In general, a closed form solution of \eqref{eq:DerMmac} cannot be found. However,
for $h=0$ and $Y=2$ we can again use the Lambert-W function to solve for $m$,
and the optimum number of blocks $m$, given a constant $R/R_0$ and the message length $K$, is
\begin{equation} \label{eq:OptM}
  m = \frac{-\ln(2)zK}
      {1 + \ln(2)\frac{zK}{X(q,2)} + W_{-1}\left( -e^{-(1+\ln(2)\frac{zK}{X(q,2)})} \right) }.
\end{equation}

To obtain the optimum number of blocks $m$ that minimizes the expected number 
of transmissions and maximizes the throughput, we solve \eqref{eq:DerMmac} and 
\eqref{eq:SolRate} jointly using numerical methods.
For $h=0$, Fig.~\ref{fig:OptM1} shows the optimum number of blocks $m$ given a
message length $K$, the number of sources $Y$, and RLNC over GF($q$).
As the total message length $K$ increases, we observe that the maximum throughput is achieved for a larger number of blocks $m$.
Since the expected coding overhead $X^*(q,Y)$ \eqref{eq:ubXq} in blocks increases with the number of sources in the star network, the optimum number of blocks $m$ increases with $Y$ for a fixed message length $K$.
On the other hand, since the expected coding overhead $X^*(q,Y)$ decreases with increasing 
Galois field size, the optimum number of blocks decreases with $q$.
For $h=16$, the optimum number of blocks is shown in Fig.~\ref{fig:MoptMACxb16}, and we see that by increasing $h$ and (using a longer header per block) the optimum number of data blocks decreases.

\subsection{Large Galois Field Considerations}

A common assumption in the analysis of network coding is that RLNC is done over a sufficiently large Galois field size that the coding overhead is negligible, i.e., 
$X^*(q,Y)\approx 0$ for large $q$.

If we set $X^*(q,Y)=0$ in \eqref{eq:ENmac}, the numerator is increasing in $m$, while the denominator is strictly decreasing in $m$. So the smallest possible $m$, i.e., $m=1$, minimizes the expected number of transmissions and maximizes throughput. Thus, in the absence of a coding overhead, i.e., for $q\rightarrow \infty$, the optimum strategy for the source is to use a channel code on the whole message and not divide it up into smaller blocks.

The same argument holds for \eqref{eq:ENtdma} and TDMA. If $X^*(q,Y)=0$, throughput is maximized if the sources choose $m=1$, i.e., the longest (and therefore strongest) possible channel code.

\section{The Broadcast Phase} \label{sec:Broad}

In this section we optimize the throughput of the broadcast phase without taking the MAC phase into account. To this end assume that the channels to the relay are error free, i.e., 
$p_\mathrm{mac}=0$. Further, since $p_\mathrm{mac}=0$, we assume that during the MAC phase the sources transmit to the relay uncoded, i.e., $R=1$, and that no header is used. A header of length $h$ is then appended to each block at the relay, and the relay uses a channel code of rate $R<1$ to protect the blocks.

\subsection{TDMA Broadcast Paired With ARQ}

We first consider the TDMA scheme, i.e., broadcast using ARQ, where every block is repeated by the relay until all the $Y-1$ sources that do not know a given transmitted message have received it correctly.

Then the expected number of blocks that the relay must broadcast is given by \cite{GTK07}
\begin{equation} \label{eq:MbroadARQ}
		{M}^{\mathrm{br}}_{\mathrm{TDMA}} = Ym \sum_{i=0}^{\infty} 1-	(1-\epsilon_{br}^i)^{Y-1},
\end{equation}
where $\epsilon_{br}$ is the block error rate of a BSC with crossover probability $p_\mathrm{br}$.
Using \eqref{eq:epsilon} and \eqref{eq:ErrorExp} we obtain for the expected number of bit transmissions by the relay
\begin{equation} \label{eq:NbroadARQ1}
\begin{aligned}
	{N}^{\mathrm{br}}_{\mathrm{TDMA}} = & k {M}^{\mathrm{br}}_{\mathrm{TDMA}} /R\\
			= & \frac{Y(K+mh)}{R} \sum_{i=0}^{\infty} 1-	
					\left( 1-2^{ -i\left( \frac{K}{m} + h\right)
				\left( \frac{R_0}{R} -1 \right)} \right)^{Y-1}.
\end{aligned}
\end{equation}
For any fixed coding rate $R$, the factor $Y(K+mh)/R$ in \eqref{eq:NbroadARQ1} as well as the block error probability $\epsilon_{br}$ are strictly increasing with increasing $m$. So the throughput for the TDMA system paired with ARQ is maximized for $m=1$ and a channel input block of size $k=K+h$.

To obtain the channel coding rate that maximizes the throughput, we transform \eqref{eq:MbroadARQ} into the finite sum
\begin{equation} \label{eq:finiteMtdmaBR}
		{M}^{\mathrm{br}}_{\mathrm{TDMA}} =  \sum_{i=1}^{Y-1} (-1)^{i+1} \binom{Y-1}{i}
				 \frac{Ym}{1-\epsilon_{br}^i},
\end{equation}
and using \eqref{eq:epsilon} and \eqref{eq:ErrorExp} we obtain
\begin{equation} \label{eq:NbroadARQ}
		{N}^{\mathrm{br}}_{\mathrm{TDMA}} =  \sum_{i=1}^{Y} (-1)^{i+1} \binom{Y}{i} 
				\frac{Y(K+mh)}{R\left( 1-2^{ -i\left( \frac{K}{m} + h\right)
				\left( \frac{R_0}{R} -1 \right)} \right)}
\end{equation}
for the expected number of bit transmissions.
We use the partial derivative of \eqref{eq:NbroadARQ} w.r.t. $R$ to obtain
\begin{equation} \label{eq:SolBroadRn}
		\sum_{i=1}^{Y} (-1)^{i} \binom{Y}{i} 
				\frac{1-2^{-izk}-ik\ln(2)\frac{R_0}{R} 2^{-iz{k}} }
				{\left( 1-2^{ -iz{k}} \right)^2} =0,
\end{equation}
where $z=(R_0/R)-1$.

\begin{figure}[t] 
    \centering
		\includegraphics[width=0.95\columnwidth]{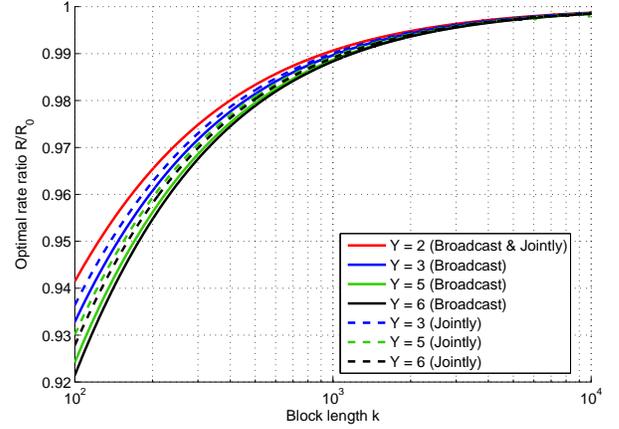}
		\caption{Optimum channel coding rate $R$ for broadcast (solid lines) and considering transmission to and from the relay jointly (dashed lines) as a fraction of the cutoff rate $R_0$ for different channel input block lengths $k$.}
		\label{fig:OptRtdma}
\end{figure}
For TDMA and $Y=2$, the channel coding rate that maximizes throughput \eqref{eq:SolBroadRn} in the broadcast phase is the same as the rate that maximizes throughput for transmission to the relay \eqref{eq:SolRate}, obtained in Section~\ref{sec:MAC}. 
In both cases, messages are transmitted from one sender to one intended receiver. 
For larger $Y$, we can numerically find the solution of \eqref{eq:SolBroadRn},
and the optimum rate ratios $R/R_0$ for broadcast from the relay for different numbers are destinations are shown as the solid lines in Fig.~\ref{fig:OptRtdma}. We see that, while \eqref{eq:SolRate} does not depend on the number of sources transmitting to the relay, during the broadcast phase the optimum channel coding rate $R$ for TDMA as a fraction of the cutoff rate $R_0$ decreases as the number of broadcast destinations increases and, for $Y>2$, is smaller than \eqref{eq:SolRate}.
The optimum number of blocks for the TDMA scheme, however, is $m=1$ for both transmission to the relay, considered in Section~\ref{sec:MAC}, and the broadcast phase.

\subsection{Broadcast Using RLNC}
\begin{figure}[t] 
    \centering
		\includegraphics[width=0.95\columnwidth]{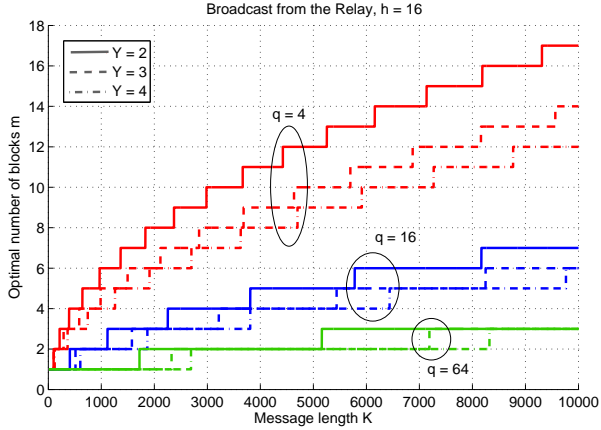}
		\vspace{-1ex}
		\caption{Optimum number of blocks for broadcast from the relay for $h=16$.}
		\label{fig:MoptBroadxb16}
		\vspace{-1ex}
\end{figure}
Using RLNC, the expected number of network coded blocks that the relay must
broadcast is given by \cite{GTK07}
\begin{equation} \label{eq:ErlncB}
	\begin{aligned}
		{M}^{\mathrm{br}}_{\mathrm{RLNC}} = & m'  + \sum_{i=m'}^{\infty} 1-
		 \left[ \sum_{j=m'}^i \left(1-\epsilon\right)^{j} \epsilon^{i-j}\right.\\
		 & \left. \binom{i}{m'}{P}(m',j-m',q)  \right]^Y,
	\end{aligned}
\end{equation}
where $m'=m(Y-1)$ is the number of unknown blocks each node $S_i$ must collect, and the probability of successful decoding given a received overhead in blocks is given by \eqref{eq:Psucc}.
The expected number of bits that the relay must transmit is then given by
\begin{equation} \label{eq:NbroadRLNC}
{N}^{\mathrm{br}}_{\mathrm{RLNC}} = \frac{k}{R}{M}^{\mathrm{br}}_{\mathrm{RLNC}}.
\end{equation}
We solve the above multidimensional optimization problem using numerical methods.
For the broadcast scenario using RLNC, Fig.~\ref{fig:MoptBroadxb16} shows the optimum number of data blocks $m$ for $h=16$. 
Comparing the optimum number of blocks in Fig.~\ref{fig:MoptBroadxb16} to the MAC phase displayed in Fig.~\ref{fig:MoptMACxb16}, the number of blocks that maximizes throughput is
generally smaller for the broadcast phase.
The most prominent difference between Fig.~\ref{fig:MoptBroadxb16} and Fig.~\ref{fig:MoptMACxb16} is that, while for the MAC phase the optimum number of data blocks increases with the number of sources, for the broadcast phase the optimum number of blocks decreases with an increase in the number of broadcast destinations $Y$, thus putting more emphasis on the channel coding being able to provide more reliable individual blocks.

\begin{figure}[t] 
    \centering
		\includegraphics[width=0.95\columnwidth]{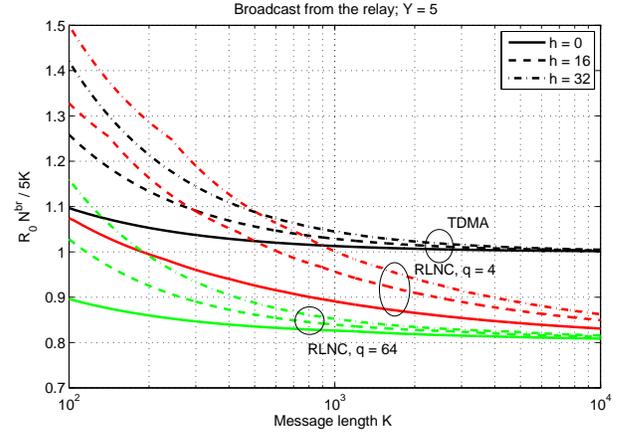}
		\vspace{-1ex}
		\caption{Expected number of broadcast transmissions per message bit multiplied by the cutoff rate $R_0$ for $Y=5$ sources.}
		\label{fig:NormNopt}
		\vspace{-3ex}
\end{figure}
Fig.~\ref{fig:NormNopt} shows the expected number of broadcast transmissions, obtained from \eqref{eq:NbroadARQ} and \eqref{eq:NbroadRLNC}, multiplied by the cutoff rate $R_0$ and divided by the total number of message bits exchanged between the $Y$ sources, $YK$.
For RLNC, the expected number of broadcast transmissions decreases with the Galois field size $q$ and increases with increasing block header size $h$. As the total message length $K$ at each source gets large, the expected number of broadcast transmissions for RLNC is smaller than for TDMA, making it more throughput-efficient, but TDMA is more throughput-efficient for small message lengths. Asymptotically, the expected number of broadcast transmissions for RLNC converges to $(Y-1)/Y$, and the convergence is faster for larger Galois fields.

\section{Joint Optimization for the Star Network} \label{sec:joint}

From the results from Sections~\ref{sec:MAC} and \ref{sec:Broad}, we see that the
number of data blocks and the channel coding rate that maximize throughput differ for
transmission from the sources to the relay and for broadcast from the relay.
In a practical system, however, it would be desirable to have the same channel coding rate and the same block size for transmission to and from the relay.\footnote{Note that, for a fixed message length of $K$ bits, keeping the number of data blocks $m$ the same is equivalent to keeping the block size constant at $K/lm$ $q$-ary symbols.}

In this Section we jointly optimize the throughput of the MAC phase investigated in Section~\ref{sec:MAC} and the broadcast phase investigated in Section~\ref{sec:Broad}. 
We assume that during the MAC phase and the broadcast phase we want to use the same number of data blocks and the same channel coding rate, so that the individual blocks are of the same size. We refer to the time it takes to transmit one block as a time slot.
For the RLNC scheme, using \eqref{eq:ErlncB}, the expected number of time slots that are occupied by transmissions in the star network is given by
\begin{equation} \label{eq:MstarRLNC}
	\begin{aligned}
			{M}^{*}_{\mathrm{RLNC}} = & 
			{M}^{\mathrm{br}}_{\mathrm{RLNC}}	\left( 1+\frac{1}{1-\epsilon_\mathrm{mac}}\right)\\
		= & \left( m'  + \sum_{i=m'}^{\infty} 1-
			  \left[ \sum_{j=m'}^i \left(1-\epsilon_{br}\right)^{j} \epsilon_{br}^{i-j}\right.\right.\\
		 	& \left.\left. \binom{i}{m'}{P}(m',j-m',q,Y)  \right]^n\right) 
		 		\left( 1+\frac{1}{1-\epsilon_{mac}}\right),
	\end{aligned}
\end{equation}
where $m'=m(Y-1)$ is the number of unknown blocks a source $S_i$ must collect and $\epsilon_{mac}$ and $\epsilon_{br}$ denote the block erasure rate during the MAC phase and the broadcast phase, respectively.
\eqref{eq:MstarRLNC} relies on the fact that, for every block that the relay broadcasts, on average $1/(1-\epsilon_\mathbf{mac})$ transmissions from the sources to the relay are necessary.

\begin{figure}[t] 
    \centering
		\includegraphics[width=0.95\columnwidth]{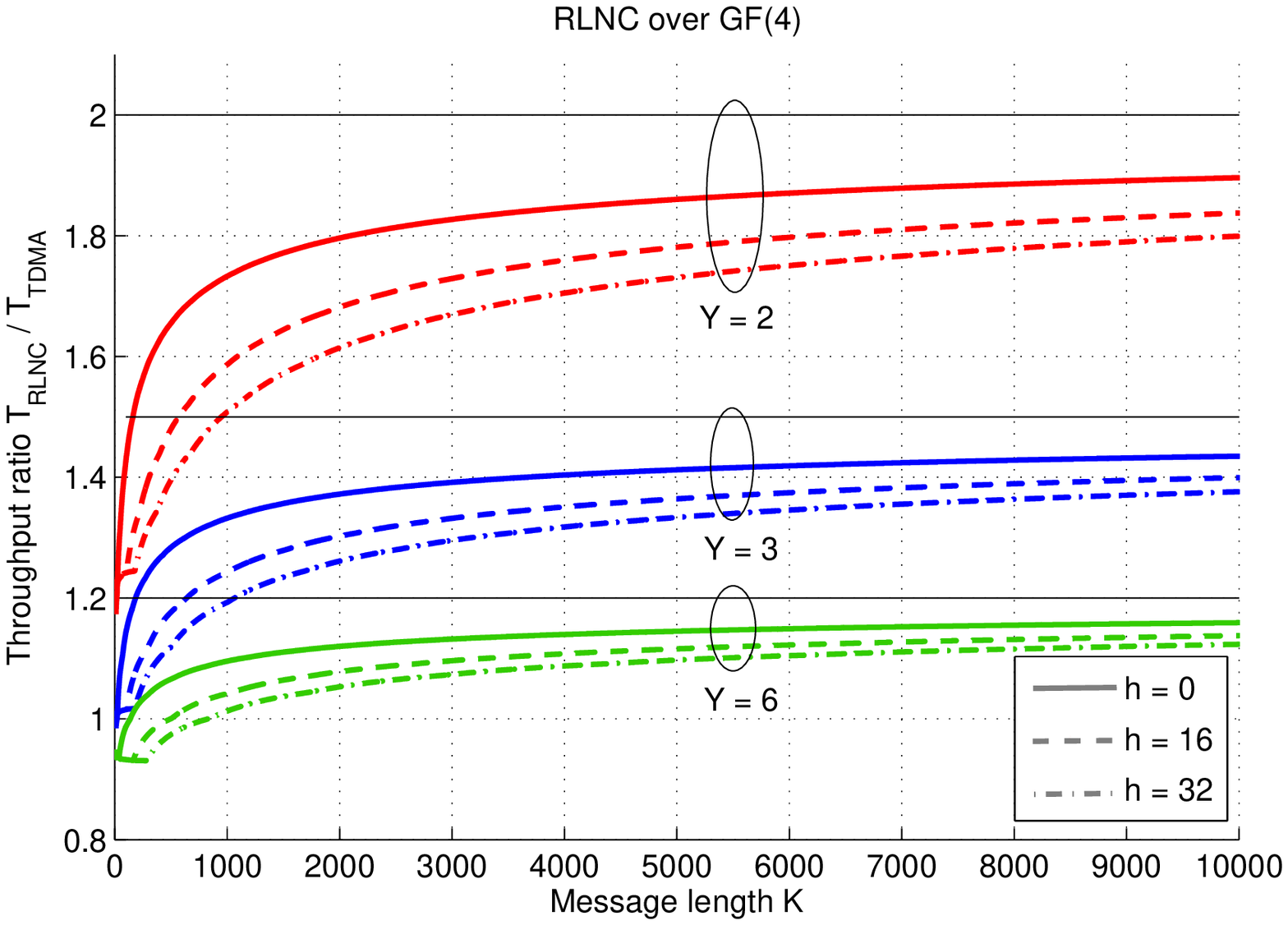}
		\caption{Average throughput ratio $T_{\mathrm{RLNC}} / T_{\mathrm{TDMA}}$ for GF(4).}
		\label{fig:TRjoint4}
\end{figure}
Similarly, using \eqref{eq:EMtdma} and \eqref{eq:finiteMtdmaBR}, for the TDMA scheme the expected number of time slots that are occupied by transmissions is given by
\begin{equation} \label{eq:MstarTDMA}
	\begin{aligned}
		{M}^{*}_{\mathrm{TDMA}} = & {M}^\mathrm{mac}_{\mathrm{TDMA}} +
						 {M}^\mathrm{br}_{\mathrm{TDMA}}\\
			 = & \frac{Ym}{1-\epsilon_\mathrm{mac}} + \sum_{i=1}^{Y-1} (-1)^{i+1} \binom{Y-1}{i}
				 \frac{Ym}{1-\epsilon_\mathrm{br}^i}.
	\end{aligned}
\end{equation}
In this case, since the throughput for both the transmission phase to the relay and the broadcast phase from the relay is maximized for $m=1$, one block of length $k=K+h$ bits for each source $S_i$ is also optimum when considering both phases jointly.

In the following, we consider the symmetric case, where $\epsilon_\mathrm{mac} = \epsilon_\mathrm{br}$, or equivalently $p_\mathrm{mac} = p_\mathrm{br}$\footnote{We assume that the sources and the relay transmit at the same power level and that therefore the channels to and from the relay have the same crossover probability.
If the transmit powers of the sources and the relay are variable, one could extend the present analysis to choose a power ratio such that, for a given channel coding rate, the same number of data blocks optimizes the throughput in both the broadcast and MAC phases.}.
In this case, the channel coding rate that maximizes the throughput for TDMA can be obtained by taking the derivative of $N^{*}_{\mathrm{TDMA}} = kM^{*}_{\mathrm{TDMA}}$ w.r.t. the channel coding rate $R$, and the optimum rate is the solution to the equation
\begin{equation} \label{eq:SolJoint}
	\begin{aligned}
		&\sum_{i=1}^{Y} (-1)^{i} \binom{Y}{i} 
				\frac{1-2^{-izk}-ik\ln(2)\frac{R_0}{R} 2^{-iz{k}} }
				{\left( 1-2^{ -iz{k}} \right)^2}\\
		& - \frac{1-2^{-zk}-ik\ln(2)\frac{R_0}{R} 2^{-z{k}} }
				{\left( 1-2^{ -z{k}} \right)^2}=0,
	\end{aligned}
\end{equation}
with $z=(R_0/R)-1$. The resulting channel coding rate that jointly maximizes throughput for the TDMA scheme star network is also depicted in Fig.~\ref{fig:OptRtdma}.

As noted in Section~\ref{sec:Broad}, the optimum channel coding rate for TDMA transmission to and from the relay, obtained separately, is the same for $Y=2$ sources. Thus, considering transmission to and from the relay jointly, the optimum rate is also given by \eqref{eq:SolRate} when $Y=2$.
For $Y>2$, the optimum channel coding rate for the star network decreases with the number of sources, similar to the TDMA broadcast case. 
However, comparing the optimum rate obtained in Section~\ref{sec:Broad} for the broadcast phase alone to the jointly optimum rate obtained from \eqref{eq:SolJoint} for the same number of sources $Y$, we find that the channel coding rate that jointly maximizes throughput for the star network is higher than the one that gives the maximum throughput for the broadcast phase alone.

\begin{figure}[t] 
    \centering
		\includegraphics[width=0.95\columnwidth]{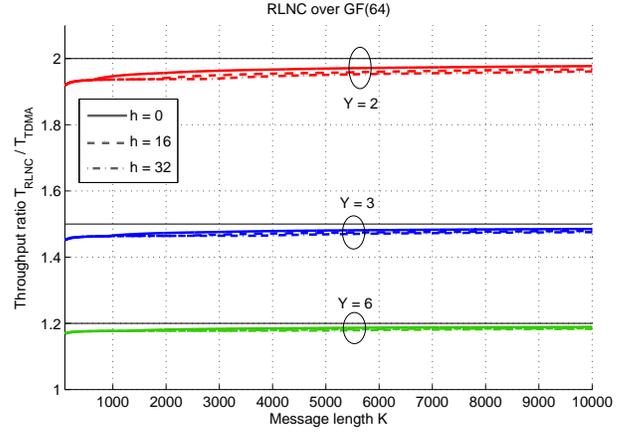}
		\caption{Average throughput ratio ${T}_{\mathrm{RLNC}} / {T}_{\mathrm{TDMA}}$ for GF(64).}
		\label{fig:TRjoint64}
\end{figure}
\begin{figure}[t] 
    \centering
		\includegraphics[width=0.95\columnwidth]{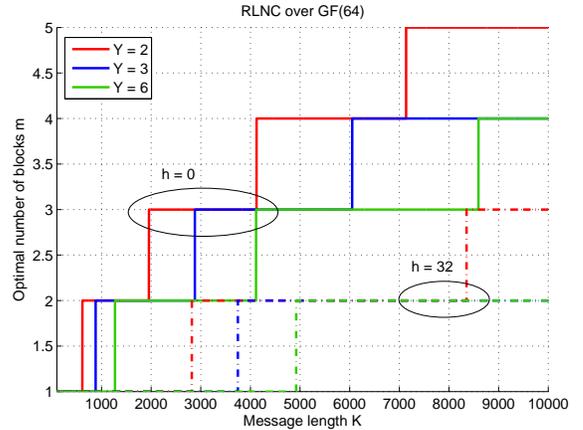}
		\caption{Optimum number of blocks $m$ in the star network for GF(64).}
		\label{fig:MoptJoint64}
\end{figure}
Figs.~\ref{fig:TRjoint4} and \ref{fig:TRjoint64} show the average throughput ratio
\begin{equation*}
	\frac{{T}_{\mathrm{RLNC}}}{{T}_{\mathrm{TDMA}}} = 
	\frac{{M}^{*}_{\mathrm{TDMA}}}{{M}^{*}_{\mathrm{RLNC}}}
\end{equation*}
of RLNC over GF(4) and GF(64) to TDMA, respectively, where \eqref{eq:MstarRLNC} and \eqref{eq:MstarTDMA} have been used to compute the ratio and the asymptotic throughput ratios are plotted as horizontal black lines.
For GF(4) and small message lengths $K$, we see in Fig.~\ref{fig:TRjoint4} that the average throughput ratio rises steeply before the curves flatten out and slowly approach their asymptotic value given by \eqref{eq:MotEx}. As the block header size $h$ increases, the average throughput ratio decreases, and a larger message length $K$ is needed to obtain a given average throughput ratio. For small message lengths $K$ and large header sizes $h$, TDMA is more throughput-efficient. For example, for RLNC over GF(4), $h=32$, and $Y=6$ sources, we require $K>900$ bits for RLNC to be more throughput-efficient than TDMA.

Employing RLNC over GF(64) which, compared to RLNC over GF(4), decreases the expected coding overhead in blocks, we see in Fig.~\ref{fig:TRjoint64} that RLNC is more throughput-efficient than TDMA for all values of $K$ and that the average throughput ratio converges much faster to its asymptotic value. For small values of $K$, there exists a region where the average throughput ratio is independent of the header size $h$. Above a certain message length $K$, however, the average throughput ratios for different header sizes $h$ separate slightly, with the region of independence extending to larger message lengths $K$ for larger values of $Y$. 
Comparing Fig.~\ref{fig:TRjoint64} to Fig.~\ref{fig:MoptJoint64}, which shows the number of data blocks that maximizes throughput for the star network with RLNC over GF(64), we see that the region where the average throughput ratio is independent of $h$ coincides with the region of $K$ values for which the optimum number of blocks is $m=1$.

\section{Bounding Channel Coding Performance Using the PPV Bound} \label{sec:Poly}

In this section we use the PPV bound given by \eqref{eq:Polyanskiy} to relate the channel coding rate $R$ and block length $n$ to the probability of block error $\epsilon$.
Compared to the random coding error exponent bound obtained by using \eqref{eq:epsilon} and \eqref{eq:ErrorExp}, where the largest possible channel coding rate is the cutoff rate, using \eqref{eq:Polyanskiy} allows channel codes that are asymptotically able to reach channel capacity.

\begin{figure}[t] 
    \centering
		\includegraphics[width=0.95\columnwidth]{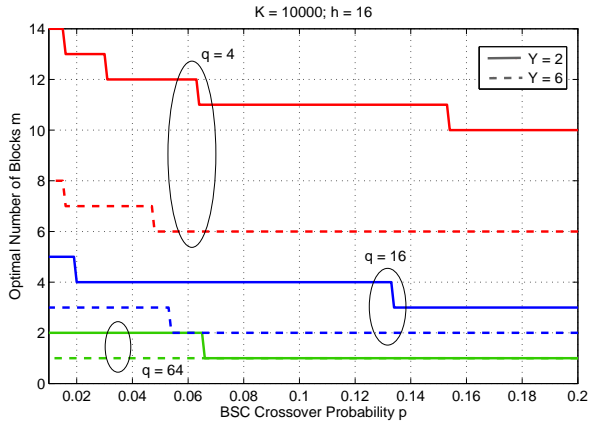}
		\caption{Optimum number of blocks $m$ in the star network vs the crossover probability $p$ of the BSC for $K=10000$ and $h=16$.}
		\label{fig:MoptPolyP}
		\vspace{-2ex}
\end{figure}
The numerical results obtained using \eqref{eq:Polyanskiy} to model channel coding in general show the same behavior as reported in Sections~\ref{sec:MAC}--\ref{sec:joint}.
We observe that the maximum throughput for RLNC is achieved for a larger number of data blocks $m$ as the message length $K$ increases and, for a fixed message length $K$, the optimum number of blocks decreases as the Galois field size of RLNC increases.
In addition, the optimum number of blocks decreases as the block header size $h$ increases in length.

\begin{figure}[t] 
    \centering
		\includegraphics[width=0.95\columnwidth]{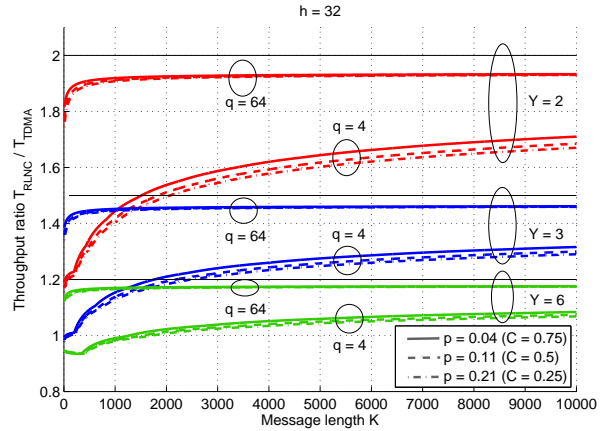}
		\caption{Throughput ratio ${T}_{\mathrm{RLNC}} / {T}_{\mathrm{TDMA}}$ for RLNC and different values of $p$.}
		\label{fig:TRjoint_Poly}
		\vspace{-2ex}
\end{figure}
In contrast to the results in Sections~\ref{sec:MAC}--\ref{sec:joint}, however, where the optimum number of data blocks $m$ did not directly depend on the crossover probability $p$ of the underlying BSC and the optimum channel coding rate could be expressed as a fraction of the cutoff rate $R_0$, using \eqref{eq:Polyanskiy} we find that the number of data blocks $m$ that maximizes throughput for RLNC varies with the crossover probability $p$.

Fig.~\ref{fig:MoptPolyP} shows the optimum number of data blocks $m$ of RLNC for star networks with $Y=2$ and $Y=6$ sources versus the BSC crossover probability $p_\mathrm{mac}=p_\mathrm{br}=p$. The message length is $K=10000$  bits, the header size is $h=16$ bits, and RLNC over GF(4), GF(16), and GF(64) is considered.

Compared to the random coding error exponent approach (see Figs.~\ref{fig:MoptBroadxb16}  and \ref{fig:MoptJoint64}), using \eqref{eq:Polyanskiy} to bound the channel coding performance generally results in a smaller optimum number of data blocks. This implies that, when maximizing throughput, the tighter bounding approach places more emphasis on channel coding and low block error probabilities and less on reducing the coding overhead of RLNC.
Furthermore, the optimum number of data blocks $m$ decreases as the channel quality 
degrades, thus requiring longer channel coded blocks, which implies more powerful codes.

The average throughput ratio ${T}_{\mathrm{RLNC}} / {T}_{\mathrm{TDMA}}$ for star networks with $Y=2$, $Y=3$, and $Y=6$ sources, RLNC over GF(4) or GF(64), and header size $h=32$ is shown in Fig.~\ref{fig:TRjoint_Poly}. 
The throughput ratios for $p_\mathrm{mac}=p_\mathrm{br}=p=0.04$, which corresponds to a BSC capacity of roughly ${C}=0.75$ bits/transmission, are plotted in solid lines, the throughput ratios for $p=0.11$ (${C}=0.5$) are plotted in dashed lines, the throughput ratios for $p=0.21$ (${C}=0.25$) are plotted in dash-dot lines, and the asymptotic throughput ratios are plotted as horizontal black lines.
We see that the average throughput ratio decreases slightly with the BSC crossover probability $p$. 
Compared to the random coding error exponent approach (see Figs.~\ref{fig:TRjoint4} and \ref{fig:TRjoint64}), using \eqref{eq:Polyanskiy} decreases the throughput ratio ${T}_{\mathrm{RLNC}} / {T}_{\mathrm{TDMA}}$, implying that employing stronger channel codes reduces the advantage that RLNC has over TDMA.

Specifically, for RLNC over GF(4), $h = 32$, $Y=6$ sources in the star network, and transmission over BSCs with $p=0.21$, we now require $K>1800$ bits for RLNC to become more throughput-efficient than TDMA.
In general the ratios in Fig.~\ref{fig:TRjoint_Poly} approach their asymptotic values
much more slowly than in Figs.~\ref{fig:TRjoint4} and \ref{fig:TRjoint64}, and we also see a larger gap between the throughput ratios employing RLNC over GF(4) and RLNC over GF(64).

\section{Conclusions}

We analyzed the joint design of channel coding on the physical layer and random linear network coding on the link layer for a star network where $Y$ outer sources send fixed length messages to each other with the help of a central relay.
For RLNC over a finite Galois field of size $q$ and messages of total length $K$ at each source, we obtain the number of data blocks and the channel coding rate that should be used to maximize the throughput of the star network using RLNC, assuming binary symmetric channels between the sources and relay and a binary adder channel model at the relay.
We also obtain the optimum number of blocks and the optimum rate for a
reference TDMA system and compare the throughputs of the two transmission schemes.
We find that, for small message lengths $K$ and RLNC over small Galois fields $q$,
TDMA is more throughput-efficient than RLNC, while RLNC is more throughput-efficient when the message length $K$ gets large.
We employ two different approaches to model the probability of channel decoding failure, a simplified random coding error exponent based on the union bound and the PPV bound for finite block lengths, where the PPV bound allows the use of more powerful, capacity achieving channel codes.
We find that the average throughput ratio of RLNC to TDMA decreases using the PPV bound, implying that for finite block lengths, stronger channel coding reduces the advantage that RLNC has over TDMA.


\end{document}